\documentclass[a4paper,UKenglish,cleveref, autoref,thm-restate]{lipics-v2021}
\usepackage{graphicx} % Required for inserting images
\usepackage{tcolorbox}
\usepackage{xcolor,colortbl,array}
\usepackage{eqnarray}
\title{Counting overlapping pairs of words} %TODO Please add

\titlerunning{Counting overlapping pairs of words} %TODO optional, please use if title is longer than one line
\author{Eric Rivals}{LIRMM, Université Montpellier, CNRS, Montpellier, France \and \url{https://www.lirmm.fr}}{rivals@lirmm.fr}{https://orcid.org/0000-0003-3791-3973}{}

\author{Pengfei Wang}{LIRMM, Université Montpellier, CNRS, Montpellier, France \and \url{https://www.lirmm.fr/~rivals/authors/pengfei-wang/}} {pengfei.wang@lirmm.fr}{https://orcid.org/0000-0001-8172-5270}{}

\authorrunning{E. Rivals and P. Wang} %TODO m, andatory. First: Use abbreviated first/middle names. Second (only in severe cases): Use first author plus 'et al.'

\Copyright{Eric Rivals and Pengfei Wang} %TODO m, andatory, please use full first names. LIPIcs license is "CC-BY";  http://creativecommons.org/licenses/by/3.0/

 %TODO m, andatory: Please choose ACM 2012 classifications from https://dl.acm.org/ccs/ccs_flat.cfm 

 %TODO m, andatory; please add comma-separated list of keywords

\category{Regular Paper} %optional, e.g. invited paper

\relatedversion{A short version of this paper is in press in the LNCS proceedings of COCOON 2025.}

\funding{E. Rivals and P. Wang are supported by the European Union’s Horizon 2020 research and innovation programme under the Marie Skłodowska-Curie grant agreement No 956229.}
\usepackage[colorinlistoftodos]{todonotes}

\newcommand{\length}[1]{\vert #1 \vert}

\newcommand{\Ss}{{{\Sigma}^{*}}}
\newcommand{\Sn}{{{\Sigma}^{n}}}

\newcommand{\Kn}{{\kappa_n}}

\newcommand{\Gn}{{\ensuremath{\Gamma_n}}}

\newcommand{\Dn}{{\ensuremath{\Delta_n}}}

\hypersetup{%
  pdfsubject="Mathematics of computing; Discrete mathematics",
  colorlinks=true,
  citecolor=blue,
  linkcolor=magenta!70!black,
  urlcolor=green!50!black,
  pdfstartpage=1}

\nolinenumbers
\begin{document}
\maketitle

\begin{abstract}
  A correlation is a binary vector that encodes all possible positions of overlaps of two words, where an overlap for an ordered pair of words $(u,v)$ occurs if a suffix of $u$ matches a prefix of $v$. As multiple pairs can have the same correlation, it is relevant to count how many pairs of words share the same correlation, depending on the alphabet size and word length $n$. We exhibit recurrences to compute the number of such pairs -- which is termed \emph{population size} -- for any correlation; for this, we exploit a relationship between overlaps of two words and self-overlap of one word. This theorem allows us to compute the number of pairs with the longest overlap of a given length, solving two open questions Gabric raised in 2022. Finally, we also provide bounds for the asymptotic population ratio of any correlation. Given the importance of word overlaps in areas like combinatorics on words, bioinformatics, and digital communication, our results may ease analyses of algorithms for string processing, code design, or genome assembly.
\end{abstract}
\newpage

\section{Introduction}

A word \(u\) overlaps a word \(v\) if a suffix of \(u\) equals a prefix of \(v\). The shared suffix-prefix is called a \emph{border} for the ordered pair of words \((u,v)\) (note that other authors call this a \emph{right border}, see~\cite{Gabric}). If \((u,v)\) has no border, it is said \emph{unbordered}. The pair \((u,v)\) is said \emph{mutually unbordered} if both \((u,v)\) and \((v,u)\) lack a border. Conversely, if both $(u,v)$ and $(v,u)$ have a border, then the pair is said to be \emph{mutually bordered}. These notions generalize the well-studied concepts of border, bordered, and unbordered words, originally defined for single words, to pairs of words.
% These notions are generalized to pairs of words, the well-studied notions of border, bordered, and unbordered words that were originally defined for single words.

% # related work paragraph on overlapping and unbordered words

Overlapping and unbordered words are central in many applications: bioinformatics, pattern matching, or code design. Computing overlaps between all pairs of sequencing reads is one step of the genome assembly task~\cite{Gusfield1997,Makinen_genome_scale_book}; several algorithms solve it in optimal time \cite{Gusfield-apsp-1992,ValimakiLM12,Tustumi-etal-apsp-2016,Lim-Park-apsp-2017}. The notion of borders is core in word combinatorics~\cite{Lothaire-COW-97,Lothaire-ACW-01}, the design of pattern matching algorithms~\cite{Knuth-Morris-Pratt-pm,Smyth-book-03}, and in the statistical analysis of pattern finding and discovery~\cite{Nielsen_expectation_search,CakirCM_conjecture_aviodpattern}. For instance, questions in vocabulary statistics deal with the distributions of the number of missing words or of common words in random texts~\cite{RahmannR00,rahmann_cpc_2003}, which depend on the overlap structure of words, and find applications in bioinformatics~\cite{Robin-etal-dna-book} or in the test of random number generators~\cite{PER:WHI:1995}.
A set of mutually unbordered words serves as code for synchronization purposes in network communication. A seminal construction algorithm appeared in 1973~\cite{Nielsen_bifix_free}, and others brought recent improvements in the design of cross bifix-free codes~\cite{Bilotta,Bajic} or non-overlapping code~\cite{Elena_et_nonoverlapping_DLT}, a topic of combinatorial interest~\cite{Blackburn_2024}.

The combinatorics of single (not pair) bordered and unbordered words over a $q$-ary alphabet has been studied in depth. For instance, the recurrence for counting the number of unbordered words of length $n$ over a q-ary alphabet was first given in~\cite{Nielsen_bifix_free}, while the recurrence for counting the number of bordered words (termed "overlapping sequences" in some articles) is proven in~\cite{Lossers_siam_1995}. From this, the probability that a random word of length $n$ is unbordered was shown to converge when $n$ tends to infinity in~\cite{Nielsen_bifix_free}, while Holub and Shallit have shown that expected maximum border length for words of length $n$ over a q-ary alphabet also converges~\cite{holub_shallit_stacs_2016}.
In a related area (but somehow more distant from our topic), other works have investigated unbordered factors of words~\cite{Holub_2012,Harju_Nowotka_stacs_2004,Loptev_cpm_2015}, a topic introduced by Ehrenfeucht and Silberger in~\cite{Ehrenfeucht_1979}.

Recently, building on ideas similar to those used in~\cite{Nielsen_bifix_free},  Gabric gave three recurrences to count bordered, mutually bordered, mutually unbordered pairs of words of length \(n\) over a \emph{k}-ary alphabet~\cite{Gabric}. In his conclusion, he raised two challenging open questions: Q1: Count the number of pairs having the longest border of length \(i\) (with \(i\) satisfying \(0 < i < n\)). Q2: What is the expected length of the longest border across all pairs of length-$n$ words? We address and solve these questions in our work (see Section~\ref{sec:solutions:gabric}).

Example: Consider the binary alphabet \(\{a, b\}\) and the following three words denoted by \(u,v,w\): \texttt{abaaa}, \texttt{aaabb}, and \texttt{abbbb}.  The pairs \((u,v)\) and \((v,w)\) both have the longest  border of length \(3\), but \((u,v)\) has \(3\) distinct non-empty  borders \texttt{aaa}, \texttt{aa}, and \texttt{a}, while \((v,w)\) has only one \texttt{abb}. The pairs $(v,u)$ and $(w,v)$ have no borders, which illustrates the asymmetry of this notion.

First, this example illustrates that the possibilities of overlap of a pair \((u,v)\) depend on the self-overlapping structure of their longest border (compare \texttt{aaa} with \texttt{abb}). Second, it shows that the self-overlap structure of the border limits the number of words having such a shared suffix-prefix, and thus the number of pairs of words to count.  Indeed, only words of length \(5\) having a suffix (resp.\ prefix) such as \texttt{aaa} or \texttt{bbb}, can participate in a pair having as many and as long borders as \((u,v)\). These observations suggest that, in response to the open question raised by Gabric, one may have to account for the complete overlap structure of a pair of words.

Other authors have proposed to encode the starting position of such overlaps in a binary vector called a \emph{correlation}~\cite{GuOd81}. In our example, the correlation of the pair \((u,v)\) is \texttt{00111}, while that of \((v,w)\) is \texttt{00100}. For any word \(z\), the correlation of \((z,z)\) is called the \emph{autocorrelation} of \(z\).  Clearly, multiple pairs can have the same correlation, and hence there are fewer correlations of length \(n\) than pairs of words of length \(n\).

Fortunately, one can build on previous studies of a set of autocorrelations, denoted \(\Gn\), and the set of correlations, denoted \(\Dn\), for all possible words of length \(n\) \cite{GuOd81,GO81b,RivalsRahmann-ICALP01,Rivals}.
It is known that the self-overlap structure of a word~\cite{GuOd81}, as well as the overlap structure of a pair of words~\cite{rivals_et_al:LIPIcs.ICALP.2023.100,Rivals_2025}, do not depend on the alphabet size (provided that the alphabet has at least two letters -- a unary alphabet makes these questions trivial). 
Combining a characterization of \(\Dn\) provided in~\cite{rivals_et_al:LIPIcs.ICALP.2023.100,Rivals_2025} and an algorithm for enumerating \(\Gn\)~\cite{C_sofsem_2025}, we can enumerate \(\Dn\) to get the list of all correlations of length \(n\).

With the terminology used in \cite{GuOd81,Rivals,rahmann_cpc_2003}, we exhibit two solutions to compute the population size of any correlation, which is the number of pairs of words having the same correlation (in Section~\ref{sec:cpopulation}). For this, we exploit two recurrences to compute the population size of autocorrelations \cite{GuOd81,RivalsRahmann-ICALP01}. With this in hand, we derive in Section~\ref{sec:solutions:gabric} a formula for the abovementioned open question 1(Corollary~\ref{cor:pop:longest:gabric}) Besides this, we provide bounds for the asymptotic behavior of the population ratio of any correlation (Theorem~\ref{thm:asymptotics:corr} Section~\ref{sec:asymptotics:ratio}), which extend the result known for autocorrelations~\cite{GuOd81}. Finally, we conclude with some open questions (Section~\ref{sec:conclusion}).  

\section{Preliminaries}\label{sec:definitions}
Let $\Sigma$ be a finite \emph{alphabet}, a set of \emph{letters} of cardinality $\sigma$. We call a sequence of elements of $\Sigma$ a \emph{string} or a \emph{word}.  The empty word is denoted by $\varepsilon$. We denote by $\Sigma^{*}$ the set of all finite words over $\Sigma$, and by $\Sigma^n$ the set of all words of length $n$ over $\Sigma$, with $n \in \mathbb{N}$.  For a word $x$, $|x|$ denotes the \emph{length} of $x$. For two words $x,y$, we denote their concatenation by $xy$, and the $k$-fold concatenation of $x$ with itself by $x^k$ for any $k >0$. For any $L \subset \Ss$, we define $x.L$ as $\{xy : y \in L\}$.

Let $u$ be a word of $\Sigma^n$. We index the letters of $u$ from $0$ to $n-1$: $u=u[0]\ldots u[n-1]$. The $i$th letter of $u$ is denoted by $u[i]$. We also denote by $u[i .. j]$ for any $0\leq i \leq j < n$ the substring of $u$ starting at position $i$ and ending at position $j$. A substring is said to be \emph{proper} iff $j-i+1 < n$.  Moreover, $u[0 .. j]$ is a prefix, $u[i .. n-1]$ is a suffix of $u$. 

%%%%%%%%%%%%%%%%%%%%%%%%%%%%%%%%%%%%%%%%%%%%%%%%%%%%%%%%%%%%%%%%%%%%%%%%%%%%%%%%%%%%%%%%%%%%%%%%%%%%%%%%%%%%%%%%%%%% 
\subsection{Definitions of borders and correlation for pairs of words}

To study overlaps between two words, we consider ordered pairs of words: we denote a pair of words $(u,v) \in \Sigma^n \times \Sigma^m$, which differs from the pair $(v,u)$.

\begin{definition}[Border of pair of words]\label{def:pairborder}
  A border of a pair of words $(u,v) \in \Sigma^n \times \Sigma^m$ is any word that is a non-empty suffix of $u$, and a non-empty prefix of $v$.  If a border exists, $(u,v)$ is said \emph{bordered}, otherwise it is \emph{unbordered}.
\end{definition}

A pair may have multiple borders, and in general, the set of borders for $(u,v)$ differs from that of $(v,u)$.
% ER explanation of differences in terminology
In his article, Gabric refers to a border of $(u,v)$ as a right border and to a border of $(v,u)$ as a left border; we use a different terminology.

Guibas \& Odlyzko~\cite{GO81b} proposed to encode in a binary vector the positions in $u$ at which a border is starting, and they named this notion the \emph{correlation} of a pair of words. From now on, for the sake of simplicity, we focus on pairs of words of equal length, denoted $n$, although our results can be generalized to the case of unequal lengths. 

\begin{definition}[Correlation]\label{def:correlation}
  Let $(u, v) \in \Sigma^n \times \Sigma^n$. The correlation of $(u,v)$, denoted by $c(u,v)$,  is a binary vector of length $n$ (i.e.,\  $c(u,v) \in \{\mathtt{0},\mathtt{1}\}^n$) satisfying $\forall i \in [0, \dots, n - 1]$
  $$ c(u,v)[i] = \begin{cases}
    \mathtt{1} & \text{ if } u[i..n-1] = v[0..n-i-1] \\
    \mathtt{0} & \text{ otherwise.}
  \end{cases}
  $$
\end{definition}

Generally  $c(u,v) \neq c(v,u)$. 
For any length $n \in \mathbb{N}$, we denote the set of all correlations for words of length $n$ by $\Delta_n$ and its cardinality by $\delta_n$ as in~\cite{Rivals}.

\begin{definition}[$\Delta_n$ and $\delta_n$]\label{def:delta}
  Let $n\in \mathbb{N}$. The set of all correlations of words of length $n$ is:
  $$ \Delta_n := \{ t \in \{\mathtt{0, 1}\}^n : \exists (u,v) \in \Sn\times\Sn : c(u,v) = t \},$$
  and its cardinality is denoted by $\delta_n$.
\end{definition}

\begin{example}
  Consider the pair of words $(u,v) = (\mathtt{aabbab},\mathtt{babbaa})$ of length $6$ over the binary alphabet $\{\mathtt{a,b}\}$. The pair $(u,v)$ has a border starting at position $3$ in $u$, and a shorter border starting at position $5$. Its correlation is $c(u,v) = \mathtt{000101}$. See Table~\ref{tab:ex:corr}.
  Of course, a permutation of the alphabet (that is, exchanging $a$ with $b$ and vice versa) yields a different pair of words, which has the same correlation as $(u,v)$. Thus, several pairs can share the same correlation.
\end{example}

\begin{table}
  \begin{center}
    \begin{tabular}{>{\columncolor[gray]{.8}}cccccccccccc>{\columncolor[gray]{.8}}c} 
      \rowcolor[gray]{.8}
      pos.  & 0 & 1 & 2 & \color{blue}{3} & 4 & \color{blue}{5} &  &  &  &  &  &
      \\ 
      $u$  & \texttt{a} & \texttt{a} & \texttt{b} & \texttt{b} & \texttt{a} & \texttt{b} & - & - & - & - & -  &$t$ \\ \hline
      $v$  & \texttt{b} & \texttt{a} & \texttt{b} & \texttt{b} & \texttt{a} & \texttt{a} & - & - & - & - &  &  $\mathtt{0}$
      \\ 
            & - & \texttt{b} & \texttt{a} & \texttt{b} & \texttt{b} & \texttt{a} & \texttt{a} & - & - & - & - &  $\mathtt{0}$
      \\
            & - & - & \texttt{b} & \texttt{a} & \texttt{b} & \texttt{b} & \texttt{a} & \texttt{a} & - & - & - &  $\mathtt{0}$
      \\
            & - & - & - & \color{blue}{\texttt{b}} & \color{blue}{\texttt{a}} & \color{blue}{\texttt{b}} & \color{blue}{\texttt{b}} & \color{blue}{\texttt{a}} & \color{blue}{\texttt{a}} & - & - &  \color{blue}{$\mathtt{1}$}
      \\
            & - & - & - & - & \texttt{b} & \texttt{a} & \texttt{a} & \texttt{b} & \texttt{a} & \texttt{b} & - &  $\mathtt{0}$
      \\
            & - & - & - & - & - & \color{blue}{\texttt{b}} & \color{blue}{\texttt{a}} & \color{blue}{\texttt{b}} & \color{blue}{\texttt{b}} & \color{blue}{\texttt{a}} & \color{blue}{\texttt{a}} &  \color{blue}{$\mathtt{1}$}
      \\
      \hline
    \end{tabular}
    \hspace{2cm}
    % table for pair (v, u) := (babbaa, aabbab) 000011
    \begin{tabular}{>{\columncolor[gray]{.8}}cccccccccccc>{\columncolor[gray]{.8}}c} 
      \rowcolor[gray]{.8}
      pos.  & 0 & 1 & 2 & 3 & \color{blue}{4} & \color{blue}{5} &   &   &   &   &   &
      \\ 
      $v$  & \texttt{b} & \texttt{a} & \texttt{b} & \texttt{b} & \texttt{a} & \texttt{a} & - & - & - & - & -  &$t$ \\ \hline
      $u$  & \texttt{a} & \texttt{a} & \texttt{b} & \texttt{b} & \texttt{a} & \texttt{b} & - & - & - & - &  &  $\mathtt{0}$
      \\ 
            & - & \texttt{a} & \texttt{a} & \texttt{b} & \texttt{b} & \texttt{a} & \texttt{b} & - & - & - & - &  $\mathtt{0}$
      \\
            & - & - & \texttt{a} & \texttt{a} & \texttt{b} & \texttt{b} & \texttt{a} & \texttt{b} & - & - & - &  $\mathtt{0}$
      \\
            & - & - & - & \texttt{a} & \texttt{a} & \texttt{b} & \texttt{b} & \texttt{a} & \texttt{b} & - & - &  $\mathtt{0}$
      \\
            & - & - & - & - & \color{blue}{\texttt{a}} & \color{blue}{\texttt{a}} & \color{blue}{\texttt{b}} & \color{blue}{\texttt{b}} & \color{blue}{\texttt{a}} & \color{blue}{\texttt{b}} & - &  \color{blue}{$\mathtt{1}$}
      \\
            & - & - & - & - & - & \color{blue}{\texttt{a}} & \color{blue}{\texttt{a}} & \color{blue}{\texttt{b}} & \color{blue}{\texttt{b}} & \color{blue}{\texttt{a}} & \color{blue}{\texttt{b}} &  \color{blue}{$\mathtt{1}$}
      \\
      \hline
    \end{tabular}
  \end{center}
  \caption{\label{tab:ex:corr} Example of correlations for the words of length $6$: $u:= \mathtt{aabbab}$ and  $v := \mathtt{babbaa}$. Left the table for $c(u,v)$: All possible shifts of $v$ to the right of $u$ are displayed on distinct lines: those at which an overlap exists are colored in blue. The last column shows $c(u,v)$ written top-down, with $\mathtt{1}$ bits colored in blue corresponding to borders. Right: same table for $c(v,u)$.}
\end{table}

%%%%%%%%%%%%%%%%%%%%%%%%%%%%%% Some def and results about periods and Gamma
A special case arises when $u$ equals $v$. Then $c(u,u)$ is called the \emph{autocorrelation} of $u$. We recall definitions of period, period set, and autocorrelation, as well as some known properties of autocorrelations that we use later on. 
% We recall definitions of period and some useful known properties of autocorrelations. 
Their proofs can be found in~\cite{rivals_et_al:LIPIcs.ICALP.2023.100,GuOd81,HALAVA,Rivals_2025}.
%%% Add two lemmas
\begin{definition}[Period]\label{def:period2}
  A word $u = u[0 .. n-1]$ has period $p \in \{0, 1, \ldots, n-1\}$ if and only if $u[0.. n-p-1] = u[p.. n-1]$, i.e.,\ for all $0 \leq i \leq n-p-1$, we have $u[i] = u[i+p]$.
\end{definition}

The zero period is called \emph{trivial}. The smallest non-trivial period of $u$ is called its \emph{basic period}. The \emph{period set} of a word $u$ is the set of all its periods and is denoted by $P(u)$. 

For all possible words of length $n \in \mathbb{N}$, the set of autocorrelations, denoted by $\Gamma_n$, is defined as:  
$\Gamma_n := \{ s \in \{\mathtt{0, 1}\}^n : \exists u \in \Sn: c(u,u) = s \}$. We denote by $\kappa_n$ the cardinality of $\Gamma_n$. Clearly, $\Gamma_n \subset \Delta_n$. When $n=0$ we consider that $\Gn = \{ \varepsilon \}$. Since this paper focuses on pairs of words, we will use $c(u,u)$ instead of $a(u)$ to represent the autocorrelation of word $u$ in the following context. 
%%%%%%%%%%%%%%%%%%%%%%%%%%%%%%%%%%%%%%%%%%%%%%%%%%%%%% 
% I change 't' to 's'

\begin{lemma}\label{lem:nested:ac}
  Let $s \in \Gn$ and $u \in \Sn$ such that $c(u,u) = s$.  Let $0 \leq p \leq q < n$ such that $s[p] = \mathtt{1}$. 
  Then, $s[q] = \mathtt{1}$ iff $u[p..n-1]$ has period $(q-p)$ (equivalently the $(q-p)$ bit in $c(u[p..n-1],u[p..n-1])$ equals $\mathtt{1}$).
\end{lemma}

\begin{lemma}\label{lem:multiply}
  Let $s\in \Gn$. For all $p$ satisfying $0\leq p < n$, and $s[p] =1$, it follows that $s[kp] = 1$ for all $k \in [2,\dots, \lfloor{\frac{n}{p}}\rfloor].$
\end{lemma}

\begin{lemma}\label{lem:ha}
  Let $\pi(u)$ be the basic period of $u \in \Sigma^n$ and $p$ be a non-trivial period. Then either $p = k \cdot \pi(u)$, $k \in [1,\dots,\lfloor{\frac{n}{\pi(u)}}\rfloor]$ or $p > n-\pi(u)$.
\end{lemma}
%%%%%%%%%%%%%%%%%%%%%%%%%%%%%%%%%%%%%%%%%%%%%%%%%%%%%%%%%% 
% ER 
\subsection{Set of all correlations of length \texorpdfstring{$n$}{Lg} and its characterization}

The first characterization of autocorrelations was given by Guibas and Odlyzko in their seminal paper~\cite{GuOd81}. They studied the cardinality of $\Gn$ and provided a lower and an upper bound for $\log(\Kn)/\log_2(n)$, and conjectured that their lower bound was also an upper bound.  They also proposed an algorithm to compute the number of words in $\Sn$ that share the same period set, which they termed the \emph{population} of an autocorrelation.  A key result of their work is the \emph{alphabet independence} of $\Gn$: Any alphabet with $\sigma > 1$ gives rise to the same set of autocorrelations, i.e., to $\Gn$.

Rivals et al.~\cite{rivals_et_al:LIPIcs.ICALP.2023.100,Rivals_2025} have characterized $\Delta_n$ and exhibited its relation to the sets $\Gamma_j$ for $0 \leq j \leq n$, which is stated below.

\begin{lemma}[Lemma 21~\cite{rivals_et_al:LIPIcs.ICALP.2023.100}]\label{lem:corr:char}
  The set of correlations of length $n$ is of the form $$\Delta_n = \left\{\mathtt{0}^{(n-j)} s, \; \text{ with }  s \in \Gamma_j \; \text{ and }\ j \in [0,\dots,n] \right\}.$$
\end{lemma}

% ER: explanation of the structure of a correlation
Lemma~\ref{lem:corr:char} gives us the \textbf{structure of any correlation} for any pair of words $(u,v)$ of length $n$: it starts with a series of $\mathtt{0}$, until the leftmost $\mathtt{0}$, which marks the position in $u$ of the longest border of pair $(u,v)$. Let $z$ denote this border and $j$ denote its length.  The above characterization is based on the fact that the suffix of length $j$ of $c(u,v)$ (the one starting with the leftmost $1$) must be the autocorrelation of $z$.  Indeed, each border of $z$ is also a border of $(u,v)$. If $j=0$, then $z$ is empty word and $c(u,v) = \mathtt{0}^n$.
Of course, if $u = v$, then the correlation of $(u,v)$ is the autocorrelation of $u$.

%%%%%%%%%%%%%%%%%%%%%%%%%%%%%% 

This characterization implies the following \textbf{partition} of $\Delta_n$:
\begin{corollary}\label{cor:delta:partition}
  $\Delta_n = \bigcup_{j=0}^{n} \{ \mathtt{0}^{n-j}s \,|\, s \in \Gamma_j \} = \bigcup_{j=0}^{n} \left(\mathtt{0}^{n-j}.\Gamma_j \right)$.
\end{corollary}

% ER rm to be put in Appendix
Since correlations (and autocorrelations) are a binary encoding of a set of positions, we can get the \emph{intersection (or union)} of two correlations by taking their logical AND (or OR). For legibility, for $t,t' \in \Delta_n$ we denote their intersection by $t \cap t'$ and their union by $t \cup t'$. We use such notation to investigate the algebraic structure of $\Delta_n$ in Appendix~\ref{sec:lattice}.

Rivals et al.~\cite{rivals_et_al:LIPIcs.ICALP.2023.100,Rivals_2025} studied the cardinalities of $\Gamma_n$ and $\Delta_n$ and proved the asymptotic convergence of ratios involving $\kappa_n$ and $\delta_n$ towards the same limit when $n$ tends to infinity. Precisely,
$\frac{\ln\kappa_n}{\ln^2(n)} \rightarrow \frac{1}{2\ln(2)},\frac{\ln\delta_n}{\ln^2(n)} \rightarrow \frac{1}{2\ln(2)}$ when $n \rightarrow \infty$.

\begin{figure}[htbp]
  \centering
  \begin{minipage}{0.75\textwidth}  % Adjust width to fit your content and page layout
    \centering
    \includegraphics[width=0.75\textwidth]{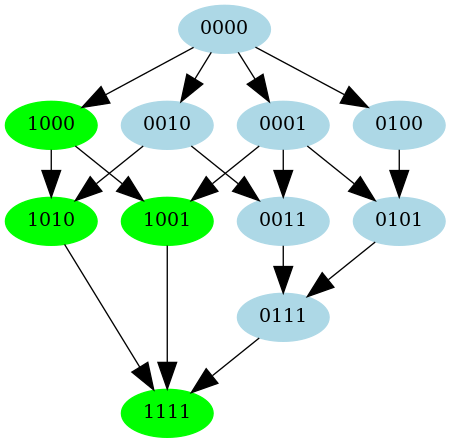}
    \caption{The lattice of \(\Delta_{4}\): each node contains a correlation as a binary vector. The elements of \(\Gamma_{4}\) are colored in green. Since the chains between $\mathtt{0000}$ and $\mathtt{1111}$ differ in length, \(\Delta_{4}\) does not satisfy the Jordan-Dedekind condition.}
    \label{fig:delta:f}
  \end{minipage}
  \\%~~~~%\hfill
  \begin{minipage}{0.745\textwidth}  % Adjust width to fit your content and page layout
    \centering
    
    \begin{tabular}{>{\columncolor[gray]{.8}}c>{\columncolor{blue}\color{white}}r>{\columncolor[gray]{.8}}r>{\columncolor{blue}\color{white}}r>{\columncolor[gray]{.8}}r}
      Correlation & \multicolumn{4}{|c|}{Population sizes} \\
      ~           & \(\; \sigma=2 \; \) & \( \;\sigma=3  \; \) & \( \;\sigma=4 \; \) & \( \;\sigma=5 \; \)\\
      \hline
      $\mathtt{0000}$ & 74 & 3678 & 45132  & 297020 \\
      $\mathtt{0001}$ & 82 & 1866 & 15108  & 74380  \\
      $\mathtt{0010}$ & 30 & 480  & 3060   & 12480  \\
      $\mathtt{0011}$ & 24 & 216  & 960    & 3000   \\
      $\mathtt{0100}$ & 16 & 162  & 768    & 2500   \\
      $\mathtt{0101}$ & 8  & 54   & 192    & 500    \\
      $\mathtt{0111}$ & 6  & 24   & 60     & 120    \\
      $\mathtt{1000}$ & 6  & 48   & 180    & 480    \\
      $\mathtt{1001}$ & 6  & 24   & 60     & 120    \\
      $\mathtt{1010}$ & 2  & 6    & 12     & 20     \\
      $\mathtt{1111}$ & 2  & 3    & 4      & 5      \\
    \end{tabular}
    \caption{Population sizes for correlations of \(\Delta_{4}\) (correlations of words of length $n=4$) and for alphabet sizes $\sigma = 2, 3, 4$ and $5$.}
    \label{tab:orgf04c32b}
  \end{minipage}
\end{figure}

Investigating the algebraic structure of $\Dn$ (See details in Appendix~\ref{sec:lattice}), we show that, alike $\Gn$, $\Delta_n$ is a lattice under set inclusion and does not satisfy the Jordan-Dedekind condition. Example~\ref{ex:delta} and Figure~\ref{fig:delta:f} illustrate the lattice structure of $\Dn$ for $n=4$.
% ER add example of Delta_4
\begin{example}\label{ex:delta}
  From Corollary~\ref{cor:delta:partition}, one has
  \(\Delta_4 = \Gamma_4 \cup ( \mathtt{0}.\Gamma_3 )\cup ( \mathtt{00}.\Gamma_2 )\cup ( \mathtt{000}.\Gamma_1 )\cup \{ \mathtt{0000} \}\). Elements of $\Gamma_4$ are shown in green background in Figure~\ref{fig:delta:f}. 
\end{example}

%%%%%%%%%%%%%%%%%%%%%%%%%%%%%%%%%%%%%%%%%%%%%%%%%%%%%%%%%%%%%%%%%%%%%% 

%%%%%%%%%%%%%%%%%%%%%%%%%%%%%%%%%%%%%%%%%%%%%%%%%%%%%%%%%%%%%%%%%%%%%% 

%%%%%%%%%%%%%%%%%%%%%%%%%%%%%%%%%%%%%%%%%%%%%%%%%%%%%%%%%%%%%%%%%%%%%%%%%%%%%%%%%%%%%%%%%%%%%%%%%%%%%%%%%%%%%%%%%%%%%%%%%%%%%%%%%%%%%%%%%%%%%%%%%%%%%%%%%%%%%%%%%%%%%%%%%%%%%%%%%%%%%%%%%%%%%%%%%%%%%%%%%%%%%%%%%%%%%%%%%%%%%%%%%%%%%%%%%%%%%%%%%%%%%%%%%%%%%%%%%%%%%%%% 

\section{Population size of a correlation}\label{sec:cpopulation}

We define \emph{population} of a correlation $t \in \Dn$ as: 
$$P(t) := \{ (u,v) \in \Sn \times \Sn \text{ such that } c(u,v) = t \}$$ 
and denote its cardinality by $p(t)$. For example, consider the correlation $t := \mathtt{01010}$ from $\Delta_5$: over the alphabet $\Sigma =\{a,b\}$, we have
$p(t) = 8$ and
\begin{equationarray*}{rcl}
  P(t) &= \{ &(\mathtt{ababa},\mathtt{babaa}), (\mathtt{ababa},\mathtt{babab}), (\mathtt{bbaba},\mathtt{babab}), (\mathtt{bbaba},\mathtt{babaa}),
  \\
  &     &(\mathtt{aabab}, \mathtt{ababa}), (\mathtt{aabab}, \mathtt{ababb}), (\mathtt{babab}, \mathtt{ababa}), (\mathtt{babab}, \mathtt{ababb}) \quad \}.
\end{equationarray*}

%%%%%%%%%%%%%%%%%%%%%%%%%%%%%%%%%%%%%%%%%%%%%%%%%%%%%%%%%%%%%%%% 
Let us give an overview of our results and detail how they generalize or improve existing ones.
% counting single population
First, for a given autocorrelation $t \in \Gn$, there exists a linear time \emph{realization} algorithm to build a binary word $u$ such that $c(u,u) = t$ \cite{2024_incremental_algo_ps}. We will exhibit such a realization algorithm for any correlation $t \in \Dn$  in Section~\ref{sec:single:pop}. In fact, this is related to counting not the pairs of $P(t)$, but single words either $u$ or $v$, for which such a pair exists.
We show a formula to determine the cardinality of the right population \( P_r(t) := \{ v \in \Sn : \exists u \in \Sn \text{ such that } c(u,v) = t \} \) or of the left population \( P_l(t) := \{ u \in \Sn : \exists v \in \Sn \text{ such that } c(u,v) = t \} \).  As \( u \) and \( v \) play a symmetrical role in \( P_l(t) \) and \( P_r(t) \), their cardinalities must be equal. We denote the cardinality of the right population of $t$ by \( p_r(t) \). Clearly, \( p^2_r(t) \) is an upper bound for \( p(t) \).

% counting pair population
In the literature, one finds two ways of computing the population size of an autocorrelation (i.e., when $t=c(u,u)$). The first recurrence formula links the population size of $t$ with that of suffix of $t$~\cite{GuOd81}[Thm 7.1]; it allows the authors to investigate the asymptotics of the population size~\cite{GuOd81}[Thm 7.2]\footnote{In their article, the authors use the term "correlation" instead of autocorrelation.}. The second recurrence formula takes advantage of the fact that $\Gn$, the set of autocorrelations of length $n$, forms a lattice with set inclusion~\cite{Rivals}.  In Section~\ref{sec:pair:pop}, we review both formulae, and exhibit two recurrence formulae for correlations: the first is based on the suffix (Theorem~\ref{thm:pop} on page~\pageref{thm:pop}), and the other is a lattice based recurrence (Theorem~\ref{thm:pop:nfc} on page~\pageref{thm:pop:nfc}).
In Appendix~\ref{sec:pop:rec:d}, we also simplify the recurrence formula from Theorem~\ref{thm:pop:corr:rec} and propose an alternative formula in Theorem~\ref{thm:pop:corr:rec:simple}.

\subsection{Realization algorithm for correlation}\label{sec:single:pop}

First, we need a simple Lemma about occurrences of a suffix of a word.
\begin{lemma}\label{lem:occ:suffix}
  Let $i>0$ and $j>0$ be two integers. Let $u \in \Sigma^{i}$ and $v \in \Sigma^{j}$.
  If the first letter of $v$ does not occur in $u$, then $v$ occurs in $uv$ only at position $i$.
\end{lemma}

Let us now state the realization problem and describe our binary realization algorithm.  
\textbf{Problem}: Consider the binary alphabet \(\Sigma = \{a, b\}\).  Let \(n>0\) and let \(t \in \Dn \). Find a pair \((u,v)\) of words over  \(\Sigma\), such that \(c(u,v) = t\).  
\\
\textbf{Algorithm}:
% autocorrelation case
If \(t[0] = \mathtt{1}\), then \(t\) is an autocorrelation. Then, call the binary realization algorithm for autocorrelation with input $t$ and return the obtained binary word~\cite{2024_incremental_algo_ps}.
% unbordered case
If \(t = \mathtt{0}^n\), the pair of words shall not overlap at all. Thus \(u := a^n\) and \(v := b^n\) satisfy the correlation vector \(t\).
% main case
Otherwise, we know there exists \(0< j < n\) and \(s \in \Gamma_j\) such that \(t = \mathtt{0}^{n-j}s\). This is the \textbf{main case}.\\
Call the binary realization algorithm for autocorrelation with \(s\) as input, and denote by \(w\) the returned binary word.  $w$ has length \(j\) and must be the suffix of $u$ and prefix of $v$.
Without loss of generality, assume \(w[0] = a\).  Then, setting \(u := b^{n-j}w\), and taking any $v$ in the set $w.\Sigma^{(n-j)}$, we get
\begin{itemize}
\item \(w\) is border of $(u,v)$, and thus \(s\) is a suffix of \(c(u,v)\);
\item $w$ has only one occurrence in $u$ by Lemma~\ref{lem:occ:suffix}, and is thus the longest border of $(u,v)$.
\end{itemize}
Hence, we get \(c(u,v) = \mathtt{0}^{n-j}s\) as required. Finally, return $(u,v)$ with $v := w.a^{n-j}$.

\medskip
From this realization algorithm, in the \textbf{main case}, we see that for a fixed $t \in \Delta_n$, once $w$ and $u$ are chosen as above, there exist $\sigma^{(n-j)}$ pairs since $v$ can be any word in $w.\Sigma^{(n-j)}$. This is a maximum for $p_r(t)$ once $w$ is fixed. Hence, we obtain the following Lemma to compute the right population size.
% ER reinsert proof here instead of in appendix
% A formal proof appears in Appendix~\ref{sec:single:pop:proof}.
\begin{lemma}\label{lem:single:pop}
  Let $t := \mathtt{0}^{n-j}s$ be in $\Dn$ with $j \in [1,\dots,n]$. Then the right population size of $t$ satisfies:
  $p_r(t) = p(s) \cdot \sigma^{(n-j)}$. 
\end{lemma}

\begin{proof}
  We give a constructive proof. Recall that $P_r(t)$ is the set of words $v$ of length $n$ whose autocorrelation admits $s$ as a prefix. 
  
  Denote $v = v_1v_2$, where $|v_1| = j, |v_2| = n-j$. Clearly, $s$ is the autocorrelation of $v_1$. The population size $p_r(t) $ equals the number of possible choices for $v_1$ times the number of possible choices for $v_2$. First, there exist $p(s)$ possible choices for $v_1$, since its autocorrelation is $s$; whereas $v_2$ can be arbitrary, which implies that the number of choices for $v_2$ is $\sigma^{n-j}$. Indeed, once a word $v$ is given, we can construct a corresponding word $u$ as follows: Denote $u = u_1v_1$ where $|u_1| = n-j$. We construct $u_1 = u[0,n-j-1]$ by choosing $u[i] \in \Sigma \setminus \{v[0]\}$, for all $i \in [0,n-j-1]$,  meaning that each letter in $u[0,n-j-1]$ differs from the first letter of $v$. It ensures that there is no overlap for $(u,v)$ before the position $n-j$.
\end{proof}

% Add the j=0 case.
Remark: if $j=0$, then the pair of words $(u,v)$ is unbordered. Note that if $vu \in \Sigma^{2n}$ with $|u|=|v|=n$ and is unbordered, then $(u,v)$ is also unbordered. Therefore, all such pairs of words (aka "bifix-free sequences") can be constructed by the algorithm of Nielsen~\cite{Nielsen_bifix_free}.

%%%%%%%%%%%%%%%%%%%%%%%%%%%%%%%%%%%%%%%%%%%%%%%%%%%%%%%%%%%%%%%%%%%%%%%%%%%%%%%% 
\subsection{Computing the population size}\label{sec:pair:pop}

Before finding a formula to compute $p(t)$, i.e., the population size, of a correlation $t$ in $\Dn$, we show that $p(t)$ is related to the population size of some autocorrelations of words of length $2n$ in Theorem~\ref{thm:number}.  To achieve this, we demonstrate two lemmas linking the borders of a pair $(u,v)$ with the borders of the word $vu$.

%%%%%%%%%%%%%%%%%%%%%%%%%%%%%%%%%%%%%%%%%%%%%%%%%%%%%%%%%%%% ALT LEMMA
\begin{lemma}\label{lem:suf}
  Let $(u,v) \in \Sigma^n \times \Sigma^n$. Then $c(u,v)$ is the suffix of length n of $c(xu, vy)$ for any words $x$, $y$ in $\Sigma^k$ for some $k \geq 0$.  
\end{lemma}
\begin{proof}
  Let $n>0$, and let $u$ and $v$ be words of $\Sigma^n$. Note that for any words $x$, $y$ in $\Sigma^k$ for some $k \geq 0$, the suffix of length $n$ of  $c(xu, vy)$ encodes the borders between the suffix of length $n$ of $xu$ (which is $u$) and the prefix of length $n$ of $vy$ (which is $v$). Thus,  $c(u,v)$ is the suffix of length n of $c(xu, vy)$.
\end{proof}

%%%%%%%%%%%%%%%%%%%%%%%%%%%%%%%%%%%%%%%%%%%%%%%%%%%%%%%%%%%% reverse lemma
\begin{lemma}\label{lem:wtouv}
  Let $w \in \Sigma^{2n}$; let $u$ and $v$ be words in $\Sn$ such that $w = vu$. If $w$ has a border, then the pair of words $(u,v)$ is bordered.
\end{lemma}
\begin{proof}
  Let $w$, $u$, and $v$ be as in the lemma.
  % case u = v
  If $u=v$, then $u$ is a border of the pair $(u,u)$.
  % General case.
  Otherwise, we have $u \neq v$.
  Let $z$ be a border of $w$. We distinguish two cases based on $\length{z}$.
  \begin{enumerate}
  \item Case 1: $|z|\in [1,\dots ,n-1]$.  Then, there exist two words $x,y$ of length $n-\length{z}$ such that $v = zy$ and $u = xz$. Thus, $z$ is a border of $(u,v)$. 
  \item Case 2: $|z| \in [n+1,\dots, 2n-1]$. Then, $w$ has a period $p := 2n-|z|$ and $p < n$ (the half $\length{w}$). According to properties of periods (Lemma~\ref{lem:multiply}), the integer $\lfloor{\frac{2n}{p}}\rfloor p$ is also period of $w$. Then, if we denote its corresponding border by $z'$, we have $\length{z'} < n$, and we are back to case 1, with $z'$ being a border of $(u,v)$.  
  \end{enumerate}
\end{proof}

%%%%%%%%%%%%%%%%%%%%%%%%%%%%%%%%%%%%%%%%%%%%%%%%%%%%%%%%%%%% MAIN THM
Before stating the theorem on the population size of a correlation, we need a notation.
Let $t \in \Dn$. We denote by $G(t)$ the set of all words of length $2n$ whose autocorrelation has $t$ as a suffix, and by $g(t)$ its cardinality. Formally, $G(t):= \{ w \in \Sigma^{2n} : t \text{ is a suffix of } c(w,w) \}$.

The following theorem shows the relation between the number of pairs of words of length $n$ and the number of specific words of length $2n$.

\begin{theorem}\label{thm:number}
  % \label{thm:pair:pop:autocorr}

  Let $t \in \Delta_n$. Then,  $p(t) = g(t)$. 
\end{theorem}
\begin{proof}
  i/ Let us first prove that $p(t) \leq g(t)$.
  Let $(u,v) \in P(t)$, that is, $c(u,v)=t$. By Lemma~\ref{lem:suf}, let $x=v$ and $y=u$, we get that $c(vu,vu)$ has $t=c(v,u)$ as a suffix, i.e.,  $vu\in G(t)$. This implies that $p(t) \leq g(t)$.
  \\
  ii/ Let us prove that $p(t) \geq g(t)$.
  Let $w \in G(t)$, and let $u$ and $v$ be words of length $n$ such that $w=vu$. Again, by Lemma~\ref{lem:suf}, we know that $c(u,v) = t$, which implies that $g(t) \leq p(t)$.

  Combining both inequalities, we get $p(t) = g(t)$, which concludes the proof. 
\end{proof}

Now we will calculate the number of pairs of words of length $n$ with the correlation $t = \mathtt{0}^{n-j}s \in \Delta_n$ where $s \in \Gamma_j$, i.e., the population size of $t$. Thanks to Theorem~\ref{thm:number}, we provide two different approaches of computing $p(t)$: the first one, based on the recurrence for the population size of autocorrelation using their recursive structure~\cite[Predicate $\Xi$]{GuOd81}, is presented in this section. %(~\ref{sec:pop:cor:nest}. 
The second approach based on the recurrence for the population size of autocorrelation that exploits the lattice structure of $\Gn$~\cite{Rivals}, is shown on page~\pageref{sec:pop:cor:la}.

\label{go:recurrence}

\subsubsection{Recurrence based on the recursive structure of autocorrelations}\label{sec:pop:cor:nest}
We review the recurrence formula given by Guibas \& Odlyzko.
Let $s \in \Gamma_j$. They define the autocorrelation of length $n$ denoted as $s_n := \mathtt{1}\mathtt{0}^{n-j-1}s$, and the sequence $\psi$ for $k \in \mathbb{Z}$ depending on $s$ as
$$\psi[k] := \begin{cases}
  0 & \text{for } k>j  \\
  s[j-k] & \text{for } 1 \leq k \leq j  \\
  \sigma^{-k} & \text{for } k < 1.
\end{cases} 
$$
We will use this definition of $s_n$ in many places. The sequence $\psi$ partitions $\mathbb{N}$ into three distinct ranges. 

For $k < 1$, $\psi[k]$ equals $\sigma^{-k}$. In the interval $1 \leq k \leq j$, $\psi[k]$ equals $1$ if $(j-k)$ is a period in $s$, and $0$ otherwise. For any $k>j$, $\psi[k]$ is consistently equal to $0$. Let $s \in \Gamma_j$ and assume fixed. Theorem~\ref{thm:autopop} states their recurrence for $p(s_n)$. 

\begin{theorem}[Population size of an autocorrelation (Theorem 7.1~\cite{GuOd81})]\label{thm:autopop}
  Let $k \in \mathbb{Z}$. Let $n,j \in \mathbb{N}$ satisfying $0 \leq j < n$. Let $s \in \Gamma_j$ and let $s_n := \mathtt{1}\mathtt{0}^{n-j-1}s$.  Then the number of words of length $n$ that have autocorrelation $s_n \in \Gn$ satisfies the recurrence:
  $$p(s_n) + \sum_{k \in \mathbb{Z}} p(s_k) \psi[2k-n] = 2\psi[2j-n]p(s), $$
  where $p(s_k) =0$ for $k<j$, and the sequence $\psi$ is defined as above. 
\end{theorem}
We state our result regarding the population size of a correlation $t=\mathtt{0}^{n-j}s$ with $s$ being fixed. See Figure~\ref{tab:orgf04c32b} for population sizes on different alphabets ($\sigma = 2,3,4,5$) for all correlations in $\Delta_4$. 
% Note that if $j=n$, then the population size of $t$ is the known population size of $s$. 
For $t\in \Gamma_n$, $p(t)$ can be directly calculated using Theorem~\ref{thm:autopop}. Therefore, in the following, we consider $t\in \Delta_n$ but exclude those in $\Gamma_n$.

\begin{theorem}[Population size of a correlation (I)] \label{thm:pop}\label{thm:pop:corr:rec}
  Let $j, n \in \mathbb{N}$ satisfying $0 \leq j < n$. Let $t := \mathtt{0}^{n-j}s$ be an element of $\Delta_n$ with $s \in \Gamma_j$. Then the population size of $t$ satisfies the recurrence
  $$p(t) = \left( \sum\limits_{\lambda = 1}^{\left\lfloor j / 2\right\rfloor} p(s_{n + \lambda}) \cdot s[j - 2\lambda]\right) + p(s_{2n}).$$
\end{theorem}
\begin{proof}

  Let $w \in G(t)$ and define the integer $\tilde{\lambda}$ as $\tilde{\lambda} := \max\{0 \leq i < n: c(w,w)[i] = \mathtt{1} \}$. According to Theorem~\ref{thm:number}, we know that $\mathit{p(t)} = g(t)$.
  Thus, we are left to show
  $$g(t) = \left( \sum\limits_{\lambda = 1}^{\left\lfloor j / 2\right\rfloor} \mathit{pop}(s_{n + \lambda}) \cdot s[j - 2\lambda]\right) + \mathit{pop}(s_{2n}).$$

  Define $s_{(2n,\tilde{\lambda})} := *^{\tilde{\lambda}} \mathtt{1}\mathtt{0}^{2n-\tilde{\lambda}-j-1}s$, where $*^{\tilde{\lambda}}$ is a short cut notation for any word in $\{0,1\}^{\tilde{\lambda}}$. We have $s_{(2n,\tilde{\lambda})}\in \{0,1\}^{2n}$. Note that this defines a binary vector of length $2n$, which may belong to $\Gamma_{2n}$ depending on the value of $\tilde{\lambda}$. Let us partition the set $G(t)$ into its subsets $P(s_{(2n,\tilde{\lambda})})$, where $s_{(2n,\tilde{\lambda})} \in \Gamma_{2n}$ 
  $$ G(t) = \bigsqcup_{\tilde{\lambda} \in [0,\dots,n-1] : s_{(2n,\tilde{\lambda})} \in \Gamma_{2n}} P(s_{(2n,\tilde{\lambda})}).
  $$

  Taking the cardinalities, for $s_{(2n,\tilde{\lambda})} \in \Gamma_{2n}$ we get 
  $$
  g(t) = \sum^{n-1}_{\tilde{\lambda}=0} p(s_{(2n,\tilde{\lambda})}) = \sum^{n-1}_{\tilde{\lambda}=1} p(s_{(2n,\tilde{\lambda})}) + p(s_{(2n,0)}).
  $$

  We distinguish different cases depending on $\tilde{\lambda}$.
  \begin{enumerate}
  \item When $\tilde{\lambda} = 0$, the autocorrelation of $w$ satisfies $c(w,w) = s_{(2n,0)} = s_{2n} = \mathtt{10}^{2n-j-1}s \in \Gamma_{2n}$. Thus the number of words $w$ having the autocorrelation $s_{(2n,0)}$ equals the population size of $s_{2n}$, i.e., $p(s_{(2n,0)}) = p(s_{2n} )$.

  \item When $\tilde{\lambda} \in [1,\dots,n-1]$, recall $s_{(2n-\tilde{\lambda})} =\mathtt{1}\mathtt{0}^{2n-\tilde{\lambda}-j-1}s$, then we have $s_{(2n,\tilde{\lambda})} = *^{\tilde{\lambda}}s_{(2n-\tilde{\lambda})} \in \{0,1\}^{2n}$. Note that not all $s_{(2n,\tilde{\lambda})}$ belongs to $\Gamma_{2n}$, but all $c(w,w)$ must have the form $s_{(2n,\tilde{\lambda})}$.
    We will identify all elements $c(w,w)$ in $\Gamma_{2n}$ that take the form $s_{(2n,\tilde{\lambda})}$.

    By the definition of $\tilde{\lambda}$, we know $\tilde{\lambda} < |w|/2$, which indicates that at most one $c(w,w)$ can possibly exist for a given $s_{(2n-\tilde{\lambda})}$. Note that $c(w,w) [2\tilde{\lambda}] = 1$ where $2\tilde{\lambda} \in [2n-j,\dots,2n-2]$, this implies $\tilde{\lambda} \geq \lceil (2n-j)/2 \rceil$.  Denote by $\pi(w)$ the basic period of $w$, then $c(w,w)$ could be decomposed as $c(w,w)= (\mathtt{10}^{\pi(w)-1})^{\alpha}s_{(2n-\tilde{\lambda})}$ where $\alpha = \tilde{\lambda}/\pi(w)$ (a proper positive divisor) by Lemma~\ref{lem:ha}.
    % indicating that $\tilde{\lambda}$ cannot be too small.   
    By Lemma~\ref{lem:nested:ac}, such an $c(w,w)$ exists precisely if $s[2\tilde{\lambda} -(2n-j)]= s[j+2\tilde{\lambda} - 2n] = 1$ since $c(w,w) [2n-j] = 1$ and $j+2\tilde{\lambda} - 2n \in [0,\dots,j-1]$. Thus we have
    $$ \sum^{n-1}_{\tilde{\lambda}=1}  p(s_{(2n,\tilde{\lambda})}) = \sum_{\tilde{\lambda} = \lceil\frac{2n-j}{2} \rceil}^{n-1}  p(s_{(2n-\tilde{\lambda} )}) s[j+2\tilde{\lambda} - 2n]. 
    $$
    By taking $\lambda = n-\tilde{\lambda}$, we obtain
    $$ \sum_{\tilde{\lambda} = \lceil\frac{2n-j}{2} \rceil}^{n-1}  p(s_{(2n-\tilde{\lambda} )}) s[j+2\tilde{\lambda} - 2n]  = \sum\limits_{\lambda = 1}^{\left\lfloor j / 2\right\rfloor} \mathit{pop}(s_{n + \lambda}) \cdot s[j - 2\lambda].$$

  \end{enumerate}
  Combine the two cases, we get
  $
  p(t) =\sum\limits_{\lambda = 1}^{\left\lfloor j / 2\right\rfloor} \mathit{pop}(s_{n + \lambda}) \cdot s[j - 2\lambda] + p(s_{2n} )$. 
\end{proof}

Observe that in Theorem~\ref{thm:pop}, calculating the population size of $t = \mathtt{0}^{n-j}s$ requires to compute $p(s_{(n+\lambda)})$  for all $\lambda \in \{0, \dots , \left\lfloor j / 2 \right\rfloor\}.$ by Theorem~\ref{thm:autopop}. Therefore, we provide a third recurrence on $t$ that calculates $p(t)$ relying only on $s$. See details in Appendix~\ref{sec:pop:rec:d}, Theorem~\ref{thm:pop:corr:rec:simple}. 

\label{rr:recurrence}

\subsubsection{Recurrence based on the lattice structure}\label{sec:pop:cor:la}

As $\Gn$ equipped with inclusion is a lattice~\cite[Theorem 3.1]{Rivals}, the successor of an autocorrelation $s$ is a more constrained autocorrelation, i.e., one that contains more periods than $s$. One can use this relationship to compute population sizes. 
From the proof of Theorem~\ref{thm:pop}, we know the autocorrelation of $w \in G(t)$ satisfies the form: $c(w,w)= (\mathtt{10}^{\pi(w)-1})^{\tilde{\lambda}/\pi(w)}s_{(2n-\tilde{\lambda})}$ where $s[j+2\tilde{\lambda} - 2n] = 1$ for all $\tilde{\lambda} \in [\lceil\frac{2n-j}{2} \rceil, \dots ,n-1]  \cup \{0\}$. Clearly $\pi(w)|\tilde{\lambda}$. Thus, we can provide another recurrence by using the notion of \emph{number of free characters (nfc for short)} introduced in~\cite{Rivals}. The \emph{nfc} of an autocorrelation $s \in \Gamma_n$ is the maximum number of positions in a word $u$ with $c(u,u)=s$ that are not determined by the periods. For instance, the \emph{nfc} of $\mathtt{100001001} \in \Gamma_9$ is 4 since a word $u$ with $c(u,u) = \mathtt{100001001}$ must satisfy \emph{character equations}: $u[0] = u[3] = u[5] = u[8], u[1]= u[6],$ and $u[2]=u[7]$. Thus $u = u[0]u[1]u[2]u[0]u[4]u[0]u[1]u[2]u[0]$ where $u[0],u[1],u[2],u[4] \in \Sigma$. Theorem~\ref{thm:pop:auto:rr} states the recurrence on population sizes. 
The nfc of an autocorrelation of length $n$ can be calculated in $\Theta(n)$ time (Algorithm 2~\cite{Rivals}).

\begin{theorem}[Population size of an autocorrelation (Theorem 6.1~\cite{Rivals})]\label{thm:pop:auto:rr}

  Let $n\in \mathbb{N}$ and $v_k$ be the $k$th ($k=1,\dots,\kappa_n$) autocorrelation of $\Gamma_n$. Let $\rho_k$ denote the number of free characters of $v_k$. The population size $p(v_k)$ satisfies the recurrence
  $$p(v_k)= \sigma^{\rho_k} -\sum_{j:v_k \subset v_j} p(v_j).
  $$
\end{theorem}
The proof of Theorem~\ref{thm:pop:auto:rr} relies on the \emph{nfc} of a given autocorrelation, on the lattice structure of $\Gamma_n$, and on the following idea.  Consider the set $\mathcal{A}$ of words that satisfy the \emph{character equations} imposed by autocorrelation $v_k$. As a word in $\mathcal{A}$ can satisfy additional character equations, $\mathcal{A}$ contains all words whose autocorrelation is $v_k$, but also words whose autocorrelations are $s_{(j)}$ with $j:v_k \subset v_j$. 
We reuse this idea to compute $p(t)$.

Let $\tilde{\lambda}_1, \ldots, \tilde{\lambda}_m$ be the proper positive divisors of $\tilde{\lambda}$.  

From the proof of Theorem~\ref{thm:pop}, the autocorrelation of $w\in G(t)$ could be decomposed based on $\tilde{\lambda}_i$ for $i\leq m$, and $\tilde{\lambda}$ for $\tilde{\lambda} \in \{\lceil\frac{2n-j}{2} \rceil, \dots ,n-1\}$ as follows:
$$c(w,w) = s_{(2n,\tilde{\lambda}, \tilde{\lambda}_i)} =  (\mathtt{10}^{\frac{\tilde{\lambda}}{\tilde{\lambda}_i}-1})^{\tilde{\lambda}_i}\mathtt{10}^{2n-\tilde{\lambda}-j-1}s.
$$

Let $s \in \Gamma_j$ and consider correlation $t := \mathtt{0}^{n-j}s$ as fixed. Recall that $g(t)$ is the cardinality of the set $G(t)$ of all words of length $2n$ whose autocorrelation has $t$ as the suffix. To state our second formula for the population size of $t$,  we assume that all autocorrelations of $\Gamma_{2n}$ have been calculated. For consistency, we use the notation $\rho_{.}$ to refer to the number of free characters of an autocorrelation (as in Theorem~\ref{thm:pop:auto:rr}). 

\begin{theorem}\label{thm:pop:nfc}
  Let $n,i \in \mathbb{N}$. Let $\rho_{(2n,\tilde{\lambda},\tilde{\lambda}_i)}$ denote the number of free characters of $s_{(2n,\tilde{\lambda}, \tilde{\lambda}_i)}$ and $\rho$ be the number of free characters of $s_{2n}$. The population size $p(t)$ satisfies:
  $$p(t) = \left( \sum_{\tilde{\lambda} = \lceil\frac{2n-j}{2} \rceil}^{n-1}  \sum_{\tilde{\lambda}_i}(\sigma^{\rho_{(2n,\tilde{\lambda},\tilde{\lambda}_i)}}- \sum_{v\in \Gamma_{2n}:s_{(2n,\tilde{\lambda}, \tilde{\lambda}_i)}\subset v} p(v))\right) +   \sigma^{\rho}- \sum_{y\in \Gamma_{2n}:s_{2n}\subset y} p(y).$$ 
\end{theorem}
\begin{proof}
  By Theorem~\ref{thm:pop} and Theorem~\ref{thm:number} we know: 
  \begin{align}
    \mathit{p(t)} = g(t) &= \sum_{\tilde{\lambda} = \lceil\frac{2n-j}{2} \rceil}^{n-1} p(s_{(2n,\tilde{\lambda})}) + p(s_{2n}) \\
                         & = \left(\sum_{\tilde{\lambda} = \lceil\frac{2n-j}{2} \rceil}^{n-1} \sum_{\tilde{\lambda}_i} p(s_{(2n,\tilde{\lambda},\tilde{\lambda}_i)}) \right)+ p(s_{2n}).\label{eq:par}
  \end{align}
  From Theorem~\ref{thm:pop:auto:rr} we get:
  \begin{align}
    p(s_{(2n,\tilde{\lambda},\tilde{\lambda}_i)} )=\sigma^{\rho_{(2n,\tilde{\lambda},\tilde{\lambda}_i)}}- \sum_{v\in \Gamma_{2n}:s_{(2n,\tilde{\lambda}, \tilde{\lambda}_i)}\subset v} p(v) \label{eq:par:lambda}
  \end{align}
  and
  \begin{align}
    p(s_{2n}) = \sigma^{\rho}- \sum_{y\in \Gamma_{2n}:s_{2n}\subset y} p(y).\label{eq:par:2n}
  \end{align}
  Combining Equations (\ref{eq:par}), (\ref{eq:par:lambda}), and (\ref{eq:par:2n}), we obtain the desired result.
\end{proof}

%%%%%%%%%%%%%%%%%%%%%%%%%%%%%%%%%%%%%%%%%%%%%%%%%%%%%%%%%%%%%%%%%%%%%% 

% 14.05.25 ER update p(z) into h(z) and \widetilde{p} into  \widetilde{h}

\section{Asymptotics on the population ratios}\label{sec:asymptotics:ratio}
The population ratio of a correlation $t \in \Dn$ is $p(t)/\sigma^{2n}$. Here, we study the asymptotic lower and upper bounds for this ratio. Before stating our result, we recall some definitions introduced by Guibas \& Odlyzko~\cite{GuOd81}.
Recall that Theorem~\ref{thm:autopop} on the population size of an autocorrelation $s_n$ relies on a sequence $\psi[k]$. They define three generating functions (with dummy variable $z$) two for $p(s_n)$ and $\psi[k]$, and introduce $\widetilde{h}(z)$, which is the normalization of ${h}(z)$ by $p(s)$. Their definitions are as follows:
$$ h(z) = \sum_{n=0}^{\infty} p(s_n) z^{-n}; \quad \psi(z) = \sum_{n=0}^{\infty}  \psi[k]z^{-n}; \quad \widetilde{h}(z) = \frac{h(z)}{p(s)}.
$$

Thus, Theorem~\ref{thm:autopop} can be rewritten as:
\begin{equation}
  \widetilde{h}(z) + \psi(z)\widetilde{h}(z^2) = 2 \psi(z)z^{-2j}.\label{eq:pop:gene:nor}   
\end{equation}
Hence, the asymptotics of $p(s_{n})$ as $n \rightarrow \infty$ with $s$ being fixed follows.
\begin{theorem}[Asymptotics on the population sizes \protect\cite{GuOd81}]\label{thm:asymptotics_auto}
  Let $\mu$ be any small positive complex number. Let $j \in \mathbb{N}$ satisfying $0 \leq j <n$.
  Let $s \in \Gamma_j$ and let $s_n := 10^{n-j-1}s$.
  The population size of $s_{n}$ divided by the population size of $s$ over an alphabet of cardinality $\sigma \geq 2$ satisfies
  $$\frac{p(s_{n})}{p(s)} = \left( \frac{2}{\sigma^{2j}} - \widetilde{h}(\sigma^2) \right) \sigma^{n} + O((\sigma + \mu)^{\frac{n}{2}}),
  $$
  where 
  $\widetilde{h}(\sigma^2)$ satisfies the Functional Equation (\ref{eq:pop:gene:nor}).

\end{theorem}

Denote $c = \frac{2}{\sigma^{2j}} - \widetilde{h}(\sigma^2)$. Note that $c$ is the asymptotic limit of $p(s_n)/(p(s) \sigma^{n})$; thus $c \cdot p(s)$ provides the limiting value of $p(s_n)/\sigma^n$.
Here, we state our result on the population size of correlation $t \in \Delta_n$ with $s$ being assumed fixed. 

In Table~\ref{tab:pop:ratio} we show, for some interesting cases,  the limiting values of $p(s_n)/\sigma^n$ and the asymptotic bounds on $p(t)/\sigma^{2n}$.

\begin{theorem}[Asymptotics on the population ratios]\label{thm:asymptotics:corr}
  Let $\mu$ be any small positive complex number.
  Let $s \in \Gamma_j$ with $j \in [0,\dots,n-1]$, and let $t := \mathtt{0}^{n-j}s$. Over an alphabet of cardinality $\sigma \geq 2$, the ratio $p(t)/p(s)$ satisfies the asymptotic inequality:
  \begin{equation}
    c \cdot \sigma^{2n} + O((\sigma + \mu)^{n})  \leq \frac{p(t)}{p(s)} < \frac{c\cdot \sigma}{\sigma -1} \cdot \sigma^{2n} + O(n (\sigma + \mu)^{n}).\label{ineq:pop:size:bounds} 
  \end{equation}
  In particular, we have the asymptotic bounds on the population ratio $p(t)/\sigma^{2n}$ 
  \begin{equation}
    c \cdot p(s) \leq \lim_{n \rightarrow \infty} \frac{p(t)}{\sigma^{2n}} < \frac{c\cdot \sigma}{\sigma -1} \cdot p(s).\label{ineq:pop:ratio:bounds}  
  \end{equation}
\end{theorem}

\begin{proof}
  By Theorem~\ref{thm:pop} on the population size of $t$, for $\lambda \in \{1, \dots , \left\lfloor j / 2 \right\rfloor\}$, we have.
  \begin{equation}
    \frac{\mathit{p(t)}}{p(s)} = \frac{\sum_{\lambda} \mathit{pop}(s_{n + \lambda}) \cdot s[j - 2\lambda]  + p(s_{2n})}{p(s)}.
    \label{eq:pop:size}
  \end{equation}
  Then Eq (\ref{eq:pop:size}) could be bounded above and below by:
  \begin{equation}
    \frac{p(s_{2n} )}{p(s)}  \leq  \frac{ \sum_{\lambda} p(s_{n + \lambda}) \cdot s[j - 2\lambda]  + p(s_{2n} )}{p(s)} < \frac{ \sum_{\lambda =0}^{n} p(s_{(n+\lambda)})}{p(s)}.
    \label{ineq:pop:size}
  \end{equation}
  From Theorem~\ref{thm:asymptotics_auto}, for any $\lambda \in [0,\dots,n]$ we have:  
  \begin{equation} \label{eq:th:asympt:go}
    \frac{p(s_{(n+\lambda)} )}{p(s)} = c \cdot \sigma^{n+\lambda} + O((\sigma + \mu)^{\frac{n+\lambda}{2}}).
  \end{equation}
  Plugging in Eq (\ref{eq:th:asympt:go}) in the left-hand side of Ineq (\ref{ineq:pop:size}) we get $$\frac{p(s_{2n} )}{p(s)} = c \cdot \sigma^{2n} + O((\sigma + \mu)^{n})$$
  and in the right-hand side of Ineq (\ref{ineq:pop:size}) we obtain:
  \begin{align*}
\sum_{\lambda =0}^{n}\frac{  p(s_{(n+\lambda)})}{p(s)}
&= &c \cdot \sum_{i=n}^{2n} \sigma^{i} +\left( O((\sigma + \mu)^{\frac{n}{2}} ) + O((\sigma + \mu)^{\frac{n+1}{2}}) + \dots + O((\sigma + \mu)^{n}) \right) \\  
&= &\frac{c\sigma}{\sigma -1} \cdot \sigma^{2n} + O(n (\sigma + \mu)^{n})  
\end{align*}
Combining both equations, we obtain Ineq (\ref{ineq:pop:size:bounds}):
$$ c \cdot \sigma^{2n} + O((\sigma + \mu)^{n})  \leq \frac{\mathit{p(t)}}{p(s)} < \frac{c\cdot \sigma}{\sigma -1} \cdot \sigma^{2n} + O(n (\sigma + \mu)^{n}). 
$$
Multiplying Ineq (\ref{ineq:pop:size:bounds}) by $p(s)/\sigma^{2n}$, we get the desired bounds Ineq (\ref{ineq:pop:ratio:bounds}) on the asymptotic behavior of the population ratio $\mathit{p(t)}/\sigma^{2n}$.

\end{proof}

%%%%%%%%%%%%%%%%%%%%%%%%%%%%%%%%%%%%%%%%%%%%%%%%%%%%%%%%%%%%%%%%%%%%%%%%%%%%%%%%
%%%%%%%%%%%%%%%%%%%%%%%%%%%%%%%%%%%%%%%%%%%%%%%%%%%%%%%%%%%%%%%%%%%%%%
\begin{table}[ht]
  \centering
  \caption{Population ratios for alphabet sizes $\sigma = 2, 3$ and $24$. Columns 3 and 4 give resp. the limiting values of $p(s_n)/\sigma^n$ and the asymptotic bounds for $p(t)/\sigma^{2n}$ for the correlations $s$ in column 2. The lower bound in col. 4 matches the value of col. 3 (taken from \cite{GuOd81} for a given $s$ and $\sigma$). The correlations $\varepsilon$ and $\mathtt{0}^{n-1}\mathtt{1}$ are the most populated ones for $\sigma=2$. For $\sigma \geq 3$, the correlation $\mathtt{0}^n$ becomes the most populated one.}\label{tab:pop:ratio}
  \begin{tabular}{>{\columncolor[gray]{.8}}c>{\columncolor{blue}\color{white}}c>{\columncolor[gray]{.8}}c>{\columncolor{blue}\color{white}}c}
    \hline
    Alphabet Size $\sigma$ & Autocorrelation $s$ & $\; p(s_n)/\sigma^n \;$ & $p(t)/\sigma^{2n}$ \\
    \hline
    2 & $\varepsilon$ & 0.268 & [0.268, 0.536) \\
    \cline{2-4}
                           & \texttt{1} & 0.300 & [0.300, 0.600) \\
    \cline{2-4}
                           & \texttt{10} & 0.110 & [0.110, 0.220) \\
    \cline{2-4}
                           & \texttt{11} & 0.089 & [0.089, 0.178) \\
    \hline
    3 & $\varepsilon$ & 0.557 & [0.557, 0.836) \\
    \cline{2-4}
                           & \texttt{1} & 0.283 & [0.283, 0.424) \\
    \cline{2-4}
                           & \texttt{10} & 0.072 & [0.072, 0.108) \\
    \cline{2-4}
                           & \texttt{11} & 0.032 & [0.032, 0.048) \\
    \hline
    24 & $\varepsilon$ & 0.957 & [0.957, 0.999) \\
    \cline{2-4}
                           & \texttt{1} & 0.042 & [0.042, 0.044) \\
    \hline
  \end{tabular}
\end{table}

%%%%%%%%%%%%%%%%%%%%%%%%%%%%%%%%%%%%%%%%%%%%%%%%%%%%%%%%%%%%%%%%%%%%%%%%%%%%%%%%
\section{Solutions to Gabric's open questions}\label{sec:solutions:gabric}

\subsection{Counting pairs of words with the longest border in a fixed range}

In the article about bordered and unbordered pairs of words~\cite{Gabric}, the author raises a challenging open question Q1: \emph{How many pairs of length-$n$ words have the longest border of fixed length $j$?}  Note that with his terminology, a border is either a right border or a left border, depending on the order of words in the pair. As the words play symmetrical roles in the definition of the border, the counts for the question are equal.

For the question, we answer a more complex question than the one asked by Gabric: \emph{How many pairs of length-$n$ words have the longest border within the fixed length range $[i..k]$}. 
From the characterization of the set of correlations (Lemma~\ref{lem:corr:char}), we know that correlations are partitioned by their longest border (Corollary~\ref{cor:delta:partition}). To consider pairs with longest border of length in the range $[i..k]$, we must count pairs having a correlation $t$ in the subset $\left\{\bigcup_{j=i}^{k} (\mathtt{0}^{n-j}.\Gamma_j)\right\}$ of $\Dn$. With the recurrence that computes the population size for any correlation $t$ (Theorem~\ref{thm:pop}), it suffices to sum up $p(t)$ overall $t$ in this subset to answer our question, which yields Theorem~\ref{thm:pop:longest:range}. By shrinking the range to a single value, we exactly answer Gabric's question, as addressed in Corollary~\ref{cor:pop:longest:gabric}.

\begin{theorem}\label{thm:pop:longest:range}
  Let $L_{[i..k]}$ be the number of pairs of words of length $n$ that have a longest border within the fixed length range $[i..k]$ where $i\leq k\in \{0,\dots,n-1\}$. Let $j\in \{i,\dots,k\}$. Let $s$ be any autocorrelation of $\Gamma_j$. Let $t: = \mathtt{0}^{n-j}s$. We have that $t \in (\mathtt{0}^{n-j}.\Gamma_j)$. Let $s_{(n+\lambda)} = \mathtt{10}^{n+\lambda-j-1}s \in \Gamma_{(n+\lambda)}$ where $\lambda \in \{0,\dots,\lfloor\frac{j}{2} \rfloor\}$. Then

\begin{align*}
  L_{[i..k]} &= \sum_{t \in (\cup_{j=i}^{k}(\mathtt{0}^{n-j}.\Gamma_j))} p(t) \\
  &= \sum_{\lambda =1}^{\lfloor\frac{j}{2}\rfloor} \sum_{s \in (\cup_{j=i}^{k}\Gamma_j)} p(s_{(n+\lambda)}) \cdot s[j-2\lambda]  + \sum_{s \in (\cup_{j=i}^{k}\Gamma_j)} p(s_{2n} ). 
\end{align*}
\end{theorem}
 
In particular, $L_{[0..k]}$ represents the number of pairs of words of length $n$ that have the longest border of length at most $k$, and $L_{[i..n-1]}$ counts the number of pairs of (distinct) words of length $n$ that have the longest border of length at least $i$. By restricting the length range to a single value $j$, we get the following corollary that answers Gabric's first question. 

\begin{corollary}\label{cor:pop:longest:gabric}
Let $L_j$ be the number of pairs of words of length $n$ that have the longest border of length $j$. Let $s$ be any autocorrelation of $\Gamma_j$. Let $t: = \mathtt{0}^{n-j}s$. Thus, we have $t  \in (\mathtt{0}^{n-j}.\Gamma_j)$.  Let $s_{(n+\lambda)} = \mathtt{10}^{n+\lambda-j-1}s \in \Gamma_{(n+\lambda)}$ where $\lambda \in \{0,\dots,\lfloor\frac{j}{2} \rfloor\}$. Then
$$
L_j = \sum_{t \in (\mathtt{0}^{n-j}.\Gamma_j)} p(t) = \sum_{\lambda =1}^{\lfloor\frac{j}{2}\rfloor} \sum_{s \in \Gamma_j} p(s_{(n+\lambda)}) \cdot s[j-2\lambda]   + \sum_{s \in \Gamma_j} p(s_{2n} ). 
$$
\end{corollary}

\subsection{Expected value of the longest border of a pair of words}\label{sec:expected}
In~\cite{Gabric}, Gabric considers a fixed alphabet size $\sigma$ and a Bernoulli i.i.d model for random words. In this model, the probability that a character occurs at any position is independent of other positions and equals $1/\sigma$.  For a fixed word length $n$, the probability of any pair of words $(u,v)$ both of length $n$ is $1/\sigma^{2n}$.
Gabric shows that the expected length of the \textbf{shortest border} of a pair of words converges to a constant. In this section, we show that the expected length of the \textbf{longest border} of a pair of words also converges, and thereby answer Q2.

Define $X$ to be the length of the longest border of a pair of words $(u,v)$. Then, the expectation of $X$ is
% ER changed to equation
\begin{equation}\label{eq:length:border:exp}
E(X) = \sum_{j=0}^{n-1} j \cdot Pr(X=j)= \sum_{j=1}^{n-1} j \cdot \frac{L_j}{\sigma^{2n}}  =\sum_{j=1}^{n-1} j \cdot \frac{ \sum_{t \in (\mathtt{0}^{n-j}.\Gamma_j)} p(t) }{\sigma^{2n}}.     
\end{equation}

\begin{theorem}\label{thm:border:exp:lg}
The asymptotic expected length of the longest border of a pair of words $(u,v) \in \Sigma^n \times \Sigma^n$ converges.
Furthermore, we have that
$$
\sum_{j=1}^{J-1}\sum\limits_{s \in \Gamma_j} \frac{ j \cdot p(s_{2n})}{\sigma^{2n}} +O(\frac{1}{\sigma^J}) \leq E_{\infty}(X) \leq \frac{\sigma}{(\sigma -1)^2},
$$
where $J \geq 2$ and $J$ is any $j$ that satisfies Theorem~\ref{thm:asymptotics_auto}.
\end{theorem}

\begin{proof}
Let $Y$ denote the length of a border of a pair of words $(u,v)$. Note that the event that \emph{the longest border of a pair of words $(u,v)$ has the length $j$} is a subset of the event that \emph{there exists a border of length $j$ of a pair of words $(u,v)$}. Hence $Pr(X=j) \leq Pr(Y=j)$ for each $j$.
We have
$$E(X) = \sum_{j=1}^{n-1} j \cdot Pr(X=j) = \sum_{j=1}^{n-1} j \cdot \frac{L_j}{\sigma^{2n}}  \leq  \sum_{j=1}^{n-1} \frac{j}{\sigma^j} = \sum_{j=1}^{n-1} j \cdot Pr(Y=j) = E(Y).$$
Hence
\begin{align}\label{ineq:up}
    E_{\infty}(X) \leq E_{\infty}(Y) =\lim_{n \rightarrow \infty} \sum_{j=1}^{n-1} \frac{j}{\sigma^j} \leq \frac{\sigma}{(\sigma -1)^2}.
\end{align}

It follows that $E_{\infty}(X)$ converges by the classic comparison test of series. Furthermore, $E_{\infty}(X)$ is bounded above by $\frac{\sigma}{(\sigma -1)^2}$.
Now we will show the lower bound. 
By Theorem~\ref{thm:pop}, $\sum_{t \in (\mathtt{0}^{n-j}.\Gamma_j)} p(t) /\sigma^{2n}$ satisfies the following inequality

\begin{align}
\frac{ \sum_{t \in (\mathtt{0}^{n-j}.\Gamma_j)} p(t) }{\sigma^{2n}} 
&= \frac{\left(\sum\limits_{\lambda = 1}^{\left\lfloor j / 2\right\rfloor} \mathit{pop}(s_{n + \lambda}) \cdot s[j - 2\lambda]\right) + \mathit{pop}(s_{2n})}{\sigma^{2n}} \notag \\
&\geq  \frac{\sum_{s \in \Gamma_j} p(s_{2n} )}{\sigma^{2n}}.\label{ineq:pop:exp}
\end{align}
Therefore, we have 
$$
E_{\infty} (X) \geq  \lim_{n \rightarrow \infty} \sum_{j=1}^{n-1} \sum\limits_{s \in \Gamma_j} \frac{ j \cdot p(s_{2n})}{\sigma^{2n}}.
$$
Recall Theorem~\ref{thm:asymptotics_auto} tells us that for fixed $j (=O(1))$, $ p(s_{2n})/\sigma^{2n}$ tends to a constant when $n \rightarrow \infty.$ Let us define $J$ as any valid $j$ that satisfies the condition of Theorem~\ref{thm:asymptotics_auto}.
Thus

\begin{align}\label{eq:poprario}
  \sum_{j=1}^{n-1}\sum\limits_{s \in \Gamma_j} \frac{ j \cdot p(s_{2n})}{\sigma^{2n}} =\sum_{j=1}^{J-1}\sum\limits_{s \in \Gamma_j} \frac{ j \cdot p(s_{2n})}{\sigma^{2n}} + \sum_{j=J}^{n-1}\sum\limits_{s \in \Gamma_j} \frac{ j \cdot p(s_{2n})}{\sigma^{2n}}.  
\end{align}

The first term in Eq.~\ref{eq:poprario} could be calculated directly by Theorem~\ref{thm:asymptotics_auto}. For the second term, $j$ ranges between $[J,n-1]$, note that there exists a $j$ (denoted by $\max$) such that when $j \in [J,\max]$, $j$ satisfy Theorem~\ref{thm:asymptotics_auto} whereas when $j \in [\max +1, n-1]$, $j$ doesn't satisfy. However, we show that the overall sum (the second term) is bounded above by $O(\frac{1}{\sigma^J})$ asymptotically. 

For any $s \in \Gamma_j$, by lemma~\ref{lem:ha}, 
the words, whose autocorrelations equal $s_{2n}$, are determined by their length-$(2n-j)$ border, i.e., $p(s_{2n}) \leq \sigma^{2n-j}$. It follows that
\begin{align}\label{ineq:any}
\frac{ p(s_{2n})}{\sigma^{2n}} \leq \frac{ \sigma^{2n-j} }{\sigma^{2n}} = \frac{1}{\sigma^j}.
\end{align}

Substitute Ineq.~\ref{ineq:any} into the second sum in Eq.~\ref{eq:poprario}, we have
$$\sum_{j=J}^{n-1}\sum\limits_{s \in \Gamma_j} \frac{ j \cdot p(s_{2n})}{\sigma^{2n}} \leq \sum_{j=J}^{n-1} \frac{j \cdot \#\Gamma_j }{\sigma^j},
$$
where $\#\Gamma_j$ is the cardinality of $\Gamma_j$. From Theorem 3.1 in \cite{rivals_et_al:LIPIcs.ICALP.2023.100}, we know that 
$$
\#\Gamma_j \leq \exp\left( \frac{3\ln j}{2} + \frac{(\ln j)^2}{2\ln 2}\right) \text{, for } j \geq 1.
$$
Meaning that 
$$\sum_{j=J}^{n-1}\sum\limits_{s \in \Gamma_j} \frac{ j \cdot p(s_{2n})}{\sigma^{2n}} \leq \sum_{j=J}^{n-1} \frac{j \cdot \exp\left( \frac{3\ln j}{2} + \frac{(\ln j)^2}{2\ln 2}\right) }{\sigma^j}.
$$
Note that $j \cdot \exp\left( \frac{3\ln j}{2} + \frac{(\ln j)^2}{2\ln 2}\right) = o(\sigma^j)$ as $ j \rightarrow \infty$, we get 
\begin{align}\label{eq:lim}
    \lim_{n \rightarrow \infty} \sum_{j=J}^{n-1} \frac{j \cdot \exp\left( \frac{3\ln j}{2} + \frac{(\ln j)^2}{2\ln 2}\right) }{\sigma^j} = O(\frac{1}{\sigma^J}).
\end{align}

Combine Eq.~\ref{ineq:up} and replace Eq.~\ref{eq:lim} in Eq.~\ref{eq:poprario}, we get bounds for our asymptotically expected length
$$
\sum_{j=1}^{J-1}\sum\limits_{s \in \Gamma_j} \frac{ j \cdot p(s_{2n})}{\sigma^{2n}} +O(\frac{1}{\sigma^J})  \leq E_{\infty}(X) \leq \frac{\sigma}{(\sigma -1)^2}.
$$

\end{proof}

%%%%%%%%%%%%%%%%%%%%%%%%%%%%%%%%%%%%%%%%%%%%%%%%%%%%%%%%%%%%%%%%%%%%%%%%%%%%%%%%
\section{Conclusion}\label{sec:conclusion}

Our work focuses on counting ordered pairs of words $(u,v)$ that satisfy a given correlation, which is a binary vector that encodes all overlaps of $u$ over $v$. The set of such pairs is called the population of the correlation, and their number the population size.  The main results on population size are stated in Theorems~\ref{thm:pop} and \ref{thm:pop:nfc}.
With these at hands, one can count the number of pairs of length-$n$ words with the longest border of length $j$ as asked by Gabric (his open question Q1)~\cite{Gabric}, or the expected length of this longest border across all pairs of length-$n$ words (his open question Q2), since the longest border is encoded in the correlation. Thus, the answer to open question Q1 is, for instance, a corollary of Theorem~\ref{thm:pop:longest:range}, which answers a more complex question. This emphasizes the importance of accounting for the complete overlap structure of a pair of words when investigating such questions. Another result illustrating this is the asymptotic convergence of the expected length of the longest border (Gabric's open question Q2--see Theorem~\ref{thm:border:exp:lg}), which is in line with the case of single words of length $n$~\cite{holub_shallit_stacs_2016}.\\
We conclude our work by proposing one conjecture and one open question:
\begin{enumerate}
    \item We conjecture that the population ratio $p(t)/\sigma^{2n}$ converges, and its asymptotic behavior equals the limiting value of  $p(s_n)/\sigma^{n}$: 
$\lim_{n\rightarrow \infty} p(t)/\sigma^{2n} = \lim_{n\rightarrow \infty} p(s_n)/\sigma^{n}$.
\item What is the variance or distribution of the length of the longest border of a pair of words?
\end{enumerate}

\bibliographystyle{plainurl}
%\bibliography{pop_size_long} 

\newpage
\appendix
%%%%%%%%%%%%%%%%%%%%%%%%%%%%%%%%%%%%%%%%%%%%%%%%%%%%%%%%%%%%%%%%%%%%%%%%%%%% 
\section{Structure of \texorpdfstring{$\Delta_n$}{Delta\_n}}
\label{sec:lattice}

In this section, we show that $\Delta_n$ is a lattice under set inclusion, and that it does not satisfy the Jordan-Dedekind condition. The Jordan-Dedekind condition requires that all maximal chains between the same two elements have the same length. This extends to $\Delta_n$ the findings of Rivals \& Rahmann~\cite{Rivals} who proved similar results for $\Gamma_n$.

First, let us now show that $\Delta_n$ is closed by intersection, for any $n > 0$. 
\begin{lemma}\label{lem:corr:intersection}
  Let $t$ and $t' \in \Delta_n$. Then $(t \cap t') \in \Delta_n$.
\end{lemma}
% ER revoir proof of this Lemma
\begin{proof}
  Let $t,t'\in \Delta_n$. By Lemma~\ref{lem:corr:char}, we can write $t = \mathtt{0}^{n-i}s_i, t' =\mathtt{0}^{n-j}s_j, s_i \in \Gamma_i, s_j \in \Gamma_j, i,j\in [0,\dots,n-1]$. We claim that if $(s_i \cap s_j) \in \Gamma_{\min(i.j)}$, then $(t \cap t') \in \Delta_n$. We distinguish two cases: 
  If $\mathbf{i = j}$, then $(s_i \cap s_j) \in \Gamma_j$ by Lemma 3.3 from \cite{Rivals}. Thus $\mathtt{0}^{n-i}(s_i \cap s_j) \in \left(\mathtt{0}^{n-j}.\Gamma_j \right) \subset \Delta_n$. 

  Otherwise, $\mathbf{i \neq j}$, and without loss of generality, we suppose $i<j$.  Let word $U \in \Sigma^i$, and word $V \in \Sigma^j$ such that $c(U,U) = s_i, c(V,V) = s_j$. Denote $V= V_1V_2$ where $|V_1| = i, |V_2| = j-i$. Let $W = (\Sigma \times \Sigma)^{i}$ such that $W[k] = (U[k], V[k]), k \in [0,i-1]$. It follows that $c(w,w) \in \Gamma_i$ (by Lemma 3.3 \cite{Rivals}). Note that $c(V,V) = c(V_1,V_1) \cup \mathtt{0}^{i}c(V_2,V_2)$. Then we have
  $(s_i \cap s_j) = c(U,U) \cap c(V,V) = c(U,U) \cap \left( c(V_1,V_1) \cup \mathtt{0}^{i}c(V_2,V_2) \right)= (c(U,U) \cap c(V_1,V_1) \cup (c(U,U) \cap \mathtt{0}^{i}c(V_2,V_2)) = c(w,w) \cup \emptyset  = c(w,w) \in \Gamma_i$. Therefore $\mathtt{0}^{n-i}(s_i \cap s_j) \in \left(\mathtt{0}^{n-i}.\Gamma_i \right) \subset \Delta_n$.  
\end{proof}

\begin{theorem}\label{thm:lattice}
  $(\Delta_n, \subset)$ is a lattice.
\end{theorem}
\begin{proof}
  Note that $(\Delta_n, \subset)$ has null element $\mathtt{0}^{n}$, and universal element $\mathtt{1}^{n}$. By Lemma~\ref{lem:corr:intersection}, $\Delta_n$ is closed under intersection. The meet $x \wedge y$ of $x,y$ is their intersection, the join $x \vee y$ of $x,y$ is the intersection of all elements containing $x,y$. The universal element ensures this intersection is not empty. 
\end{proof}

% ER write about chains and JD condition.
By Lemma~\ref{lem:corr:char}, we have $\Gamma_n$ is strictly included in $\Delta_n$. As any autocorrelation has its leftmost bit equal to $\mathtt{1}$, and only the autocorrelations have this property in $\Delta_n$, it follows that only an autocorrelation can be a successor of an autocorrelation. Moreover, $\mathtt{1}\mathtt{O}^{n-1}$ is a successor of the null element $\mathtt{O}^{n}$. It follows that, between the null and universal element of $\Delta_n$, there is a chain of length strictly smaller than $n$ that goes through a chain between $\mathtt{1}\mathtt{O}^{n-1}$ and the universal element $\mathtt{1}^{n}$ and traverses only nodes that are autocorrelations, by Lemma 3.5 from \cite{Rivals} when $n>6$. More exactly this chain has length $\lfloor n/2 \rfloor + 1$. 

For any $n>2$, the following chain $\mathtt{O}^{n} \prec \mathtt{O}^{n-1}\mathtt{1} \prec \ldots \prec \mathtt{O}^{n-i}\mathtt{1}^{i}$ (with $i$ in ${2, \ldots, n-2}$) to  $\mathtt{O}\mathtt{1}^{n-1}$, and finally to the universal element $\mathtt{1}^{n}$ exists in $\Delta_n$. This chain is maximal and has length $n$ -- which is the maximal length of a chain in $\Delta_n$.
Since there exist two maximal chains of different length between the null and universal elements of $\Delta_n$, when $n>6$, and visual inspection of $\Delta_4$ and $\Delta_5$ confirms the same property, we obtain this Theorem.

\begin{theorem}\label{thm:JD}
  For $n>3$, the lattice $\Delta_n$ does not satisfy the Jordan-Dedekind condition.
\end{theorem}

%%%%%%%%%%%%%%%%%%%%%%%%%%%%%%%%%%%%%%%%%%%%%%%%%%%%%%%%%%%% 
\section{Population size of a correlation: recurrence II}
\label{sec:pop:rec:d}
\begin{theorem}\label{thm:pop:corr:rec:simple}\label{thm:pop2}
  Let $k, \lambda, j, n \in \mathbb{N}$ satisfying $0 \leq \lambda, j < n$. Let $s \in \Gamma_j$ be a fixed element. Define $t := \mathtt{0}^{n-j}s$ to be an element of $\Delta_n$. Then, $p(t)$, the population size of $t$ satisfies the recurrence
  % $$p(t) = \sum_{\lambda} (2\psi[2j+\lambda-2n]p(s) -  \sum_{k} p(s_k) \psi[2k-2n+\lambda])  \cdot s[j+2\lambda - 2n] + p(s_{2n} ) 
  % $$
  \begin{align*}
    p(t) &=  \sum_{\lambda = \lceil\frac{2n-j}{2} \rceil}^{n-1} 2 p(s) \psi[2j+\lambda-2n] \; s[j+2\lambda - 2n] \\
           & -  \sum_{\lambda = \lceil\frac{2n-j}{2} \rceil}^{n-1} \sum_{k} p(s_k) \psi[2k-2n+\lambda]) \; s[j+2\lambda - 2n] + p(s_{2n} ),     
  \end{align*}
  where $p(s_k) = 0$ for $k<j$, and $\psi$ is defined as above.
\end{theorem}
\begin{proof}
  We just substitute the recurrence on $s_{(2n-\lambda)}$ by Theorem~\ref{thm:autopop} to Theorem~\ref{thm:pop}
  $$p(s_{(2n-\lambda)}) = 2p(s) \psi[2j+\lambda-2n]  - \sum_{k} p(s_k) \psi[2k-2n+\lambda].
  $$
  
\end{proof}

\end{document}